\documentclass[RNAAS]{aastex63}

\newcommand{\arf}[1]{\texttt{arf}}
\newcommand{\rmf}[1]{\texttt{rmf}}

\received{June 1, 2019}
\revised{January 10, 2019}
\accepted{\today}
\submitjournal{RNAAS}

\shorttitle{}
\shortauthors{Rhea et al.}
\usepackage{amsmath}
\newcommand{\oii}{[{\sc O\,ii}]}
\newcommand{\oiii}{[{\sc O\,iii}]}
\newcommand{\nii}{[{\sc N\,ii}]}
\newcommand{\sii}{[{\sc S\,ii}]}

\graphicspath{{./}{figures/}}
\usepackage{comment}
\usepackage{physics}
\begin{document}

\title{LUCI: A Python package for SITELLE spectral analysis}

\correspondingauthor{Carter Lee Rhea}
\email{carterrhea@astro.umontreal.ca}

\author[0000-0003-2001-1076]{Carter Rhea}
\affiliation{Département de Physique, Université de Montréal, Succ. Centre-Ville, Montréal, Québec, H3C 3J7, Canada}
\affiliation{Centre de recherche en astrophysique du Québec (CRAQ)}

\author[0000-0001-7271-7340]{Julie Hlavacek-Larrondo}
\affiliation{Département de Physique, Université de Montréal, Succ. Centre-Ville, Montréal, Québec, H3C 3J7, Canada}

\author[0000-0002-5136-6673]{Laurie Rousseau-Nepton}
\affiliation{Canada-France-Hawaii Telescope, Kamuela, HI, United
States}

\author[0000-0002-2478-5119]{Benjamin Vigneron}
\affiliation{Département de Physique, Université de Montréal, Succ. Centre-Ville, Montréal, Québec, H3C 3J7, Canada}

\author[0000-0002-8323-7809]{Louis-Simon Guité}
\affiliation{Département de Physique, Université de Montréal, Succ. Centre-Ville, Montréal, Québec, H3C 3J7, Canada}

\begin{abstract}
High-resolution optical integral field units (IFUs) are rapidly expanding our knowledge
of extragalactic emission nebulae in galaxies and galaxy clusters. By studying the spectra
of these objects -- which include classic HII regions, supernova remnants, planetary nebulae,
and cluster filaments -- we are able to constrain their kinematics (velocity and velocity dispersion).
In conjunction with additional tools, such as the BPT diagram (e.g. \citealt{baldwin_classification_1981}; \citealt{kewley_host_2006}), we can further classify
emission regions based on strong emission-line flux ratios. \texttt{LUCI} is a simple-to-use python module
intended to facilitate the rapid analysis of IFU spectra. \texttt{LUCI} does this by integrating
well-developed pre-existing python tools such as \texttt{astropy} and \texttt{scipy} with new
machine learning tools for spectral analysis (\citealt{rhea_machine_2020}). Furthermore, \texttt{LUCI} provides
several easy-to-use tools to access and fit SITELLE data cubes.
\end{abstract}

\keywords{SITELLE data reduction, IFU, Machine Learning}

\section{Statement of Need}

Recent advances in the science and technology of integral field units (IFUs) have resulted in the creation of the high-resolution, wide field-of-view (11 arcmin x 11 arcmin) instrument SITELLE (\citealt{drissen_sitelle_2019}) at the Canada-France-Hawaii Telescope. Due to the large field-of-view and the small angular resolution of the pixels (0.32 arcseconds), the resulting data cubes contain over 4 million spectra. Therefore, a simple, fast, and adaptable fitting code is paramount -- it is with this in mind that we created \texttt{LUCI}.

\section{Functionality}
At her heart, like any fitting software, \texttt{LUCI} is a collection of pre-processing and post-processing functions coupled with a fitting algorithm (in this case, a \texttt{scipy.optimize.minimize} function call) to extract information from a spectrum. That being said, \texttt{LUCI} is more than a simple wrapper for a fitting call.

Since SITELLE data cubes are available as \texttt{HDF5} files, \texttt{LUCI} was built to parse the original file and create an instance of a \texttt{LUCI} cube which contains the 2D header information and a 3D NumPy array (spatial X, spatial Y, spectral). Once the data cube has been successfully converted to a \texttt{LUCI} cube, there are several options for fitting different regions of the cube (e.g., \texttt{fit\_cube}, \texttt{fit\_entire\_cube}, \texttt{fit\_region}) or fitting single spectra (e.g., \texttt{fit\_spectrum\_region}). The primary use case of \texttt{LUCI} is to fit a region of a cube defined either as a box (in this case, the user would employ the \texttt{fit\_cube} method and pass the limits of the bounding box) or to fit a region of the cube defined by a standard \texttt{ds9} file (in this case, the user would pass the name of the region file to \texttt{fit\_region}). Regardless of the region being fit, the user needs only specify the lines they wish to fit, the fitting function, and the constraint relations between the lines. We currently support all standard lines in the SN1 filter (\oii3626 \& \oii3629), SN2 filter (\oii4959, \oiii5007, \& H$\beta$), and SN3 filter (\sii6716, \sii6731, \nii6548, \nii6583, \& H$\alpha$). The user chooses between three fitting functions: a pure Gaussian, and pure sinc function, or a sinc function convolved with a Gaussian (\citealt{martin_optimal_2016}). In either case, \texttt{LUCI} will solve for the three primary quantities of interest, which are the \textbf{amplitude} of the line, the \textbf{position} of the line (often described as the velocity and quoted in km/s), and the \textbf{broadening} of the line (often described as the velocity dispersion and quoted in units of km/s). \texttt{LUCI}'s suite of post-processing tools allows us to extract kinematic parameters and line fluxes directly from these quantities.

\subsection{Fitting Functions}
The three fitting functions are mathematically described below where $p_0$ corresponds to the \textbf{amplitude}, $p_1$ corresponds to the line \textbf{position}, and $p_2$ corresponds of the \textbf{broadening} of the line.

The pure Gaussian function is expressed as
\begin{equation}
    f(x) = p_0\exp{\frac{-(x-p1)^2}{(2*p_2^2)}}
\end{equation}

The pure sinc function is expressed as
\begin{equation}
    f(x) = p_0{\frac{\text{sin}\big((x-p_1)/p_2\big)}{(x-p_1)/p_2}}
\end{equation}

The convolved sincgauss function is expressed as
\begin{equation}
    f(x) = p_0\exp{-b^2}\Bigg(\frac{\text{erf}(a-ib)+\text{erf}(a+ib)}{2\text{erf}(a)}\Bigg)
\end{equation}

where $x$ represents a given spectral channel, $a = \frac{p_2}{(\sqrt{2}\sigma)}$, $b=\frac{x-p_1}{\sqrt{2}\sigma}$, where $\sigma$ is the
pre-defined width of the sinc function. We define this as $\sigma = \frac{1}{2\text{MPD}}$ where \textbf{MPD} is the maximum path difference. We note that erf is the error function.

\subsection{Kinematic Equations}
In each case, after solving for these values, the velocity and velocity dispersion (also called the broadening) are calculated using the following equations:

\begin{equation}
    v [\text{km/s}] = 3e5*\Bigg(\frac{p_1' - v_0}{v_0}\Bigg)
\end{equation}
where $3e5$ represents the speed of light in kilometers per second, $p_1'$ is $p_1$ in nanometers, and $v_0$ is the reference wavelength of the line in nanometers.
\begin{equation}
    \sigma [\text{km/s}] = 3e5*\Bigg(\frac{p_2}{p_1}\Bigg)
\end{equation}
where again $3e5$ represents the speed of light in kilometers per second.

\subsection{Flux Equations}

Similarly, we define the flux for each fitting function.

Flux for a Gaussian Function:
\begin{equation}
    \text{Flux} [\text{erg/s/cm}^2] = \sqrt{2\pi}p_0p_2
\end{equation}

Flux for a sinc Function:
\begin{equation}
    \text{Flux} [\text{erg/s/cm}^2] = \pi p_0p_2
\end{equation}

Flux for a sincgauss Function:
\begin{equation}
    \text{Flux} [\text{erg/s/cm}^2] = p_0\frac{\sqrt{2\pi}p_2}{erf(\frac{p_2}{\sqrt{2}\sigma})}
\end{equation}

\hfill

\subsection{Uncertainty Estimates}
A full Bayesian approach is implemented in order to determine uncertainties on the three key
fitting parameters ($p_0, p_1,$ and $p_2$) using the \texttt{emcee} python package (\citealt{foreman-mackey_emcee_2013}). It can be activated by added \texttt{bayes\_bool=True} to the argument of any fitting function. Thus, we are able to calculate posterior distributions for each parameter. There is a dedicated discussion to this in our online documentation.
Additionally, \texttt{LUCI} can calculate the uncertainties by calculating the covariance matrix of the fit; this allows users to calculate uncertainties using a classical method by adding the argument \texttt{uncertainty\_bool=True} to any fitting function (again, there is a dedicated discussion on this calculation in our online documentation).

\section{Discussion}
\subsection{Installation, Documentation, \& Examples}
\texttt{LUCI} is available as an open-source software package on our GitHub page: \url{https://github.com/crhea93/LUCI}. Moreover, we have included, on this page, a detailed set of instructions to install and use the software package. We have included detailed documentation outlining the features of \texttt{LUCI} and the mathematics behind her at \url{https://crhea93.github.io/LUCI/index.html}. On this site, we have included several key examples that can be adapted to any use-case of the software.

\subsection{Other Software}
Several fitting software packages exist for fitting generalized functions to optical spectra (such as \texttt{astropy}; \citealt{collaboration_astropy_2013}; \citealt{collaboration_astropy_2018}). Additionally, there exist software for fitting IFU datacubes for several instruments such as MUSE (\citealt{richard_reduction_2012}) and SITELLE (\citealt{martin_orbs_2012}). Although these are mature codes, we opted to write our own fitting package that is transparent to users and highly customize-able.

\acknowledgments

The authors would like to thank the Canada-France-Hawaii Telescope (CFHT), which is operated by the National Research Council (NRC) of Canada, the Institut National des Sciences de l'Univers of the Centre National de la Recherche Scientifique (CNRS) of France, and the University of Hawaii. The observations at the CFHT were performed with care and respect from the summit of Maunakea, which is a significant cultural and historic site.

C. L. R. acknowledges financial support from the physics department of the Universit\'{e} de Montr\'{e}al, IVADO, and le fonds de recherche -- Nature et Technologie.

J. H.-L. acknowledges support from NSERC via the Discovery grant program, as well as the Canada Research Chair program.

\software{astropy \citep{collaboration_astropy_2013, collaboration_astropy_2018}, scipy \citep{virtanen_scipy_2020}, numpy \citep{harris_array_2020}, matplotlib \citep{hunter_matplotlib_2007}, ipython \citep{perez_ipython_2007}, tensorflow \citep{abadi_tensorflow_2015}, keras \citep{chollet_keras_2015}, emcee \citep{foreman-mackey_emcee_2013}}

\bibliography{Luci}{}
\bibliographystyle{aasjournal}

\end{document}